\documentclass[prb,twocolumn,amsmath,amssymb,aps]{revtex4-1}
\usepackage[utf8]{inputenc}
\usepackage{feynmp}
\usepackage{graphicx}
\usepackage{adjustbox}
\usepackage{dcolumn}
\usepackage{bm}
\usepackage{comment}
\usepackage{hyperref}
\usepackage{mathtools}
\usepackage{float}
\hypersetup{
     colorlinks   = true,
     linkcolor    = blue,
     citecolor    = blue
}
\usepackage[all]{hypcap}
\newcommand{\beq}{\begin{equation}}
\newcommand{\eeq}{\end{equation}}
\newcommand{\ba}{\begin{array}{ccc}}
\newcommand{\ea}{\end{array}}

\newcommand{\tqh}{{\textrm {qh}}}
\newcommand{\tsc}{{\textrm {sc}}}

\def\bea{\begin{eqnarray}}
\def\eea{\end{eqnarray}}

\begin{document}

\title{Chiral supercurrent through a quantum Hall weak link}
\author{Yahya Alavirad$^1$}\thanks{These two authors contributed equally to this work}
\author{Junhyun Lee$^1$}\thanks{These two authors contributed equally to this work}
\author{Ze-Xun Lin$^{2,3}$}
\author{Jay D. Sau$^1$}

\affiliation{
$^1$Department of Physics, Condensed Matter theory center and the Joint Quantum Institute, University of Maryland, College Park, MD 20742, USA}
\affiliation{
$^2$Department of Physics, University of Texas at Austin, Austin, Texas 78712, USA}
\affiliation{
$^3$National Laboratory of Solid State Microstructures and Department of Physics, Nanjing University, Nanjing 210093, China}

\date{\today}

\begin{abstract}
We use a microscopic model to calculate properties of the supercurrent carried by chiral edge states of a quantum Hall weak link. This ``chiral" supercurrent is qualitatively distinct from the usual Josephson supercurrent in that it cannot be mediated by a single edge alone, i.e., both right and left going edges  are needed. Moreover, chiral supercurrent was previously shown to obey an unusual current-phase relation with period $2 \phi_0=h/e$, which is twice as large as the period of conventional Josephson junctions. We show that the ``chiral" nature of this supercurrent is sharply defined, and is robust to interactions to infinite order in perturbation theory. We compare our results with recent experimental findings of Amet {\it et al.}~\cite{Amet966} and find that quantitative agreement in magnitude of the supercurrent can be attained by making reasonable but critical assumptions about the superconductor quantum Hall interface.
\end{abstract}

\maketitle

\section{Introduction} Recently it has been recognized that proximity induced coupling between edge state of a quantum Hall (QH) system and a superconductor (SC) provides a rich playground to observe novel and exotic phenomena. In particular, these systems were theoretically demonstrated to support Majorana and parafermionic zero modes~\cite{Clarke13,Linder12,Cheng12,Maissam,YA16}. Additionally, SC/QH/SC Josephson junctions can allow for a new type of supercurrent carried by the chiral edge states~\cite{Ma93,Ostaay11,stone,Fisher,PhysRevLett.84.1804}. This ``chiral" supercurrent is qualitatively distinct from the usual Josephson supercurrent in that it cannot be mediated by a single edge alone, i.e., both right and left moving edges need to be involved. Such chiral supercurrents obey an unusual current-phase relation with the period $2 \phi_0=h/e$, which is twice as large as the period of conventional Josephson junctions~\cite{Ostaay11}. Josephson currents in related systems have also been studied in Refs.~\cite{hm1,hm2,hm3,hm4,hm5,hm6}.

Interestingly, in the past few years several different experiments have succeeded in creating a QH/SC interface~\cite{Amet966,Lee17,Wan15,1801.01447,park2017propagation}. In particular, Amet {\it et al.}~\cite{Amet966} found convincing evidence of chiral supercurrents carried by the quantum Hall edge states. In the semiclassical limit, the chiral supercurrents are propagated by quasiparticles bound in skipping orbits that are undergoing Andreev reflection at the SC interface. Such quasiparticles are expected to be slow such that this supercurrent might be too weak to be observed, however a theoretical understanding of the magnitude of the chiral supercurrent is lacking. Additionally, in apparent contradiction with theory~\cite{Ostaay11,p1,p2}, the experiment observed usual $\phi_0=h/2e$ periodicity for the current-phase relation, which would arise from tunneling through a conventional (non-chiral) insulator.
\begin{figure}[t]
\centering
	\vspace{1mm}
\includegraphics[width=0.75\columnwidth,keepaspectratio]{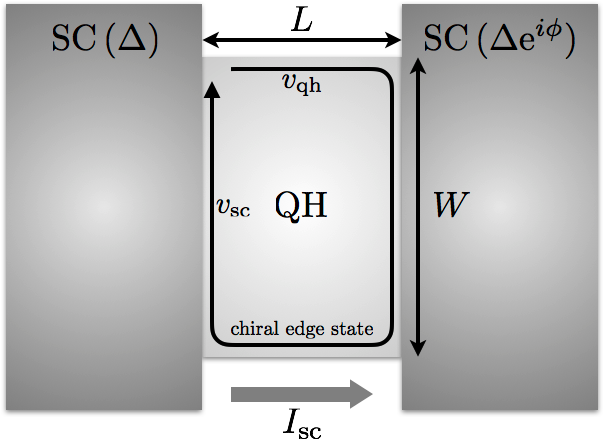} 
\caption{Top view of the system, comprised of a quantum Hall weak link attached to a pair of $s$-wave superconductors with a phase difference $\phi$. Edge velocity $v_{\tqh}$ is renormalized to $v_{\tsc}$ along the superconducting contacts. $I_\tsc$ is the chiral suppercurrent through the weak link.}
\label{fig:1.pdf}
\end{figure}

In this article, we use a microscopic model to calculate the supercurrent carried by chiral edge states of a spin degenerate quantum Hall weak link in a geometry that is similar to the experiments of Ref.~\onlinecite{Amet966} (see Fig.~\ref{fig:1.pdf}). We find that the obtained supercurrent, calculated for experimentally reasonable parameters, is quantitatively consistent with the measurement in Ref.~\onlinecite{Amet966}. In particular, we show that proximity induced edge velocity renormalization along the SC contacts and surface transparency (which is constrained by normal state conductance) play a crucial role in controlling the magnitude of the supercurrent. 
We then show that an ideal chiral quantum Hall edge state, even when interactions are included to all orders in perturbation theory, only carries chiral supercurrent, and claim that this can be used as a sharp definition for ``chiral" supercurrents. We are unable to explain the anomalous $\phi_0=h/e$ periodicity observed in the experiment.

\section{Model} We work within the geometrical setup depicted in Fig.~\ref{fig:1.pdf}. We use $x$ as a one dimensional coordinate for the QH boundary which is in contact with the SC at $L < x < L + W$ and $2L + W < x< 2(L + W)$.  Note that $x = 0$ is identified with $x = 2(L + W)$. Without the SCs, the continuum Hamiltonian describing the spin degenerate chiral quantum Hall edge is given by $H_{\textrm {QH}} =-i \hbar v_{\tqh} \int dx  \Psi^\dagger (x) \partial_x \Psi(x)$. Here $\Psi^\dagger(x) = (\psi^\dagger_\downarrow(x) ,~ \psi_\uparrow(x))$ is a two component spinor, $\psi^\dagger_{\downarrow/\uparrow}(x)$ is the pseudo-spin down/up Fermionic creation operator, and $v_{\tqh}$ is the QH edge velocity.

We now include the SCs and their couplings with the QH edge to $H_{\textrm {QH}}$. The full Hamiltonian describing the SC/QH/SC junction is $H_{\textrm {tot}}=H_{\textrm{QH}}+H_{\textrm{SC}}+H_{\textrm t}$. $H_{\textrm{SC}}$ is the BCS mean field Hamiltonian describing the SCs; we assume the SCs to be $s$-wave. $H_{\textrm t}$ is the Hamiltonian describing normal electron hopping between the SC and the QH edge along the superconducting interface. Note that we have not included the QH bulk states in $H_{\textrm{tot}}$ since they are gapped. 

Coupling with the SC induces a gap to the QH boundary spectrum at the interface. In the experimentally relevant limit where the superconducting gap $|\Delta_0|$ is much smaller than the cyclotron frequency $\hbar \omega_c$, this effect can be accounted for by including a self-energy $\Sigma (\omega)$ to the QH edge~\cite{Tudor10}. Following the results of Ref.~\onlinecite{Tudor10}, we can write the self-energy as:
 \begin{align}
\Sigma (\omega) \approx -\lambda \frac{\omega \tau_0+\Delta_0 \tau_x}{\sqrt{|\Delta_0|^2-\omega^2}}\label{selfe}.
\end{align}
Here $\tau$ is the Pauli matrix in the $\Psi(x)$ spinor space ($\tau_0$ is the $2\times2$ identity matrix) and $\lambda$ is a constant characterizing the SC/QH interface which increases as the coupling (hopping) between the SC and QH becomes larger. $\lambda$ is also related to the broadening of edge state's single particle spectral function caused by the coupling to the SC.

The effective Hamiltonian of the QH edge proximate to the SC ($H^{\textrm {eff}}_{\textrm{QH/SC}}$) can be defined by $(\omega-H_{\textrm {QH}}-\Sigma (\omega))^{-1}\propto (\omega-H^{\textrm {eff}}_{\textrm{QH/SC}})^{-1}$. 
In the low energy limit, $\omega\ll|\Delta_0|$, the self energy (Eq.~\eqref{selfe}) can be expanded to first order in $\omega$ and the effective Hamiltonian becomes: 
\begin{align}
	H^{\textrm {eff}}_{\textrm{QH/SC}} =  \int dx \Psi^{\dagger}(x) \left[ \frac{-i \hbar v_{\tqh}}{1+\lambda/|\Delta_0|}  \tau_0 \partial_x + \frac{\lambda \Delta_0}{\lambda+|\Delta_0|} \tau_x \right] \Psi(x).\label{vren}
\end{align}
The first term shows that the edge velocity $v_{\tqh}$ is strongly renormalized to $v_{\tsc} = v_{\tqh}/(1+\lambda/|\Delta_0|)$ in proximity to the SC. 
Within the semiclassical skipping orbit picture, this velocity renormalization can be attributed to the time delay associated with Andreev reflection from the SC surface.
In each period, a skipping electron spends an additional time of order $\hbar/\Delta_0$ in the SC, which changes the the period from $T_\tqh=\pi/\omega_c$ to $T_\tsc\approx\pi(1/\omega_c+\hbar/\Delta_0)$. 
The finite (imperfect) transparency of the interface, $|t|$, can be considered as the probability of Andreev reflection and can be taken into account by modifying $T_\tsc\approx\pi\left(1/\omega_c+|t|\hbar/\Delta_0\right)$. This leads to a renormalized edge velocity,
\begin{align}
v_{\tsc} = v_{\tqh} \left[1+\frac{|t|\hbar \omega_c}{\Delta_0}\right]^{-1}. \label{Tapx2}
\end{align}
We will use this semiclassical result to estimate the value of $\lambda$. Our subsequent calculation shows that the velocity renormalization plays a crucial role in controlling the magnitude of the chiral supercurrent.

The second term of Eq.~(\ref{vren}) describes the typical proximity induced superconductivity on a one-dimensional system. 
Note that the induced superconducting order parameter is also renormalized from its bare value by a factor of $1/(1+|\Delta_0|/\lambda)$.
However, $\lambda \gg |\Delta_0|$ in our parameter regime which is relevant to the experiment, and the effect of $\Delta_0$ renomarlization is not significant as that of the velocity. 

The final aspect to consider in our model is the phase difference between the two SCs. The superconducting phase difference $\phi$ shown in Fig.~\ref{fig:1.pdf} can be eliminated by a gauge transformation that introduces a vector potential $a(x)$ given by:
\begin{align}\label{gauge}
a(x)=\left\{\begin{array}{cl}
-\phi/2L & \text{for } 0<x<L \\
\phantom{-} \phi/2L & \text{for } L+W<x<2L+W \\
0 & \text{elsewhere } 
\end{array} \right. .
\end{align}

Combining $H_{\textrm{QH}}$ and $H^{\textrm {eff}}_{\textrm{QH/SC}}$ with the vector potential $a(x)$, we obtain the effective Hamiltonian describing the entire edge of the QH junction:
\begin{align}
	H =  \int dx \Psi^{\dagger}(x) \Big[\hbar v(x) ( -i \tau_0 \partial_x - a(x) \tau_z) +\Delta(x) \tau_x \Big] \Psi(x) . \label{Htot}
\end{align}
Here $v(x)$ and $\Delta(x)$ are the position dependent edge velocity and superconducting order parameter satisfying $v(x) = v_{\tqh}$ and $\Delta(x) = 0$ for $0<x<L$ and $L+W<x<2L+W$; $v(x)=v_{\tsc}$ and $\Delta(x) = \Delta$ elsewhere, where $\Delta$ is the induced superconducting order parameter $\Delta = \frac{\lambda}{\lambda+|\Delta_0|}\Delta_0$. 

\section{Josephson supercurrent} The supercurrent in the SC/QH/SC junction is given by the phase derivative of the free energy: $I_{\tsc} = -\frac{2e}{\hbar} \frac{\partial F}{\partial \phi}$. 
By expanding the free energy in imaginary time and accounting for our gauge choice (Eq.~\eqref{gauge}) the expression for supercurrent can be written in terms of single particle Green's functions~\cite{fnt},
\begin{align}\label{I3}
I_{\tsc}= -\frac{ev_{\tqh}}{\beta L}\sum_m   & \left[ \int_0^L dx\,  {\textrm {Tr}} \Big[ G(x,x;i\omega_m) \tau_z \Big] \right. \\ \newline \nonumber
  &- \left. \int_{L+W}^{2L+W} dx\,  {\textrm {Tr}} \Big[ G(x,x;i\omega_m) \tau_z \Big]  \right].
\end{align}
Here $ G(x,x;i\omega_m)$ is the single particle Green's function, $\omega_m=(2m+1)\pi/\beta$ is the Fermonic Matsubara frequency, and $\beta=1/k_B T$ is the inverse temperature.
Note that $G(x,x;i\omega_m)$ is singular for Hamiltonians which are first order in derivative (such as Eq.~\eqref{Htot}).
We regularize this singularity as $G(x,x;i\omega_m)=\lim_{\varepsilon\rightarrow 0} [G(x+\varepsilon,x;i\omega_m)+G(x-\varepsilon,x;i\omega_m)]/2$, however, our results are independent of the regularization scheme we choose.

To calculate the Green's function, we solve the defining differential equation $\left( i \omega_m - H \right) G(x, x' ; i \omega_m) = \delta(x - x')$. Assuming $0<x<L$, integrating this equation around the QH edge but the delta function $\delta(x - x')$ gives:
\begin{align}
\lim_{\varepsilon\rightarrow 0^+}G( x-\varepsilon,x ; i \omega_m)=M \left[ \lim_{\varepsilon\rightarrow 0^+} G( x+\varepsilon , x; i \omega_m) \right]. \label{gre1}
\end{align}
$M$ is an $x$ independent $2\times2$ matrix given by,
\begin{align}
	M = e^{-\frac{2 \omega_m}{\hbar}(\frac{L}{v_{\tqh}} + \frac{W}{v_{\tsc}})}  e^{i\mathbf{n} \cdot {\boldsymbol \tau}} ,
\end{align}
where $\mathbf{n}$ is a three-component vector depending on the parameters of the system.
Integrating the differential equation through the delta function from $x-\varepsilon$ to $x+\varepsilon$ gives the second equation: 
\begin{align}
\lim_{\varepsilon\rightarrow 0^+} \big[G( x+\varepsilon,x ; i \omega_m) -G( x-\varepsilon,x ; i \omega_m)\big]= -i / \hbar v_{\tqh}. \label{gre2}
\end{align}
Eqs.~\eqref{gre1}, \eqref{gre2} give a complete solution for the Green's function $G(x,x;i\omega_m)$ in our regularization scheme.
Together with the straightforward extention of $G(x,x;i\omega_m)$ for $L+W<x<2L+W$, we can calculate $I_\tsc$ using Eq.~\eqref{I3}.

\subsection{Chiral nature of supercurrent and its Interaction robustness}
The chiral nature of the supercurrent is manifest from Eq.~\eqref{I3}. To see this consider the case where only one the left/right going edges exist, i.e., the other edge is either obstructed or equivalently its length goes to infinity. In this limit for $\omega_m>0$, $M\rightarrow0$ which in turn shows $\lim_{\varepsilon\rightarrow 0^+}G( x-\varepsilon,x ; i \omega_m)=0$. Plugging this results back into Eqs.~\eqref{I3}, \eqref{gre2}, together with the straightforward extention to $\omega_m<0$, gives vanishing supercurrent $I_\tsc=0$. Note that the crucial condition leading to this results is $G( x-\varepsilon,x ; i \omega_m)=0$, that is, absence of backward propagation in a chiral edge. This property is the key feature distinguishing chiral and non-chiral supercurrents (e.g. in quantum spin Hall edge states\cite{hart2014induced}). 

One might wonder whether the introduction of interactions allows chiral quantum Hall edge states to carry non-chiral or conventional supercurrents through Cooper pair transport on the edge. Such non-chiral supercurrent could potentially explain the conventional supercurrent periodicity observed in the experiment~\cite{Amet966}. However, this turns out to be impossible and as we show below, a chiral quantum Hall edge state can only carry a chiral supercurrent. 
 
\begin{figure}[t]
\centering
\includegraphics[width=\columnwidth,keepaspectratio]{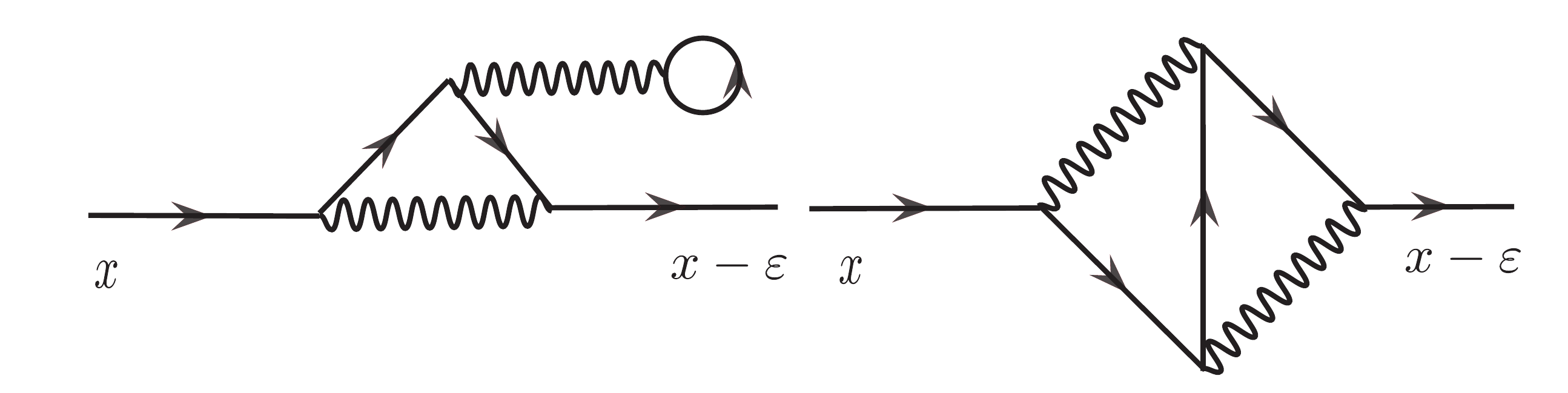} 
\caption{Typical Feynman diagrams used to calculate backward propagating interacting Green's fucntion, $\lim_{\varepsilon\rightarrow 0^+}G( x-\varepsilon,x ; i \omega_m)$. The solid lines are bare Fermionic propagators and the wiggly lines are propagators for the interaction. Note that our Feynman rule only allows a single connected string of bare Fermionic Green's function: this ensures that every diagram contributing to the backward propagating `interacting' Green's fucntion contains at least one backward propagating `bare' Green's fucntion, which leads to $\lim_{\varepsilon\rightarrow 0^+}G( x-\varepsilon,x ; i \omega_m) = 0$.\label{fig:2.pdf}}
\end{figure}

To see this, we first note that Eq.~\eqref{I3} still holds in the presence of interactions (since extra interaction terms are not flux dependent). Green's function defining equation will be modified to $\left( i \omega_m - H - \Sigma \right) G(x, x' ; i \omega_m) = \delta(x - x')$, where $\Sigma$ is the interaction induced self-energy (not to be confused with the self-energy in Eq.~\eqref{selfe}). As long as $\Sigma$ is finite we can still integrate this equation to re-obtain Eq.~\eqref{gre2}. It is then easy to see that in the absence of backwards propagation, $\lim_{\varepsilon\rightarrow 0^+}G( x-\varepsilon,x ; i \omega_m)=0$,  supercurrent still vanishes, $I_{\tsc}=0$. The limit $\lim_{\varepsilon\rightarrow 0^+}G( x-\varepsilon,x ; i \omega_m)$ can be calculated using Feynman diagrams of the type shown in Fig.~\ref{fig:2.pdf}. However, the presence of at least one backward propagating bare Fermionic Green's function in each diagram forces all terms to vanish identically, which in turn guarantees  $\lim_{\varepsilon\rightarrow 0^+}G( x-\varepsilon,x ; i \omega_m)=0$ and $I_{\tsc}=0$ to infinite order in perturbation theory.
\subsection{Explicit form of the supercurrent}
We now return to the explicit calculation of $I_{\tsc}$. Directly solving Eqs.~\eqref{gre1}, \eqref{gre2} to obtain the Green's function and using the results in Eq.~\eqref{I3} gives,
\begin{align}\label{fnl}
I_{\tsc}=-\sum_{\omega_m}&\frac{4e}{\beta\hbar}\sin \phi \, \sin^2 \left(\frac{\Delta W}{\hbar v_{\tsc}}\right) \left[ (1+\cos\phi)\cos \left(\frac{2\Delta W}{\hbar v_{\tsc}}\right)\right. \nonumber \\
&\left.+1-\cos\phi-2\cosh\left(\frac{2\omega_m}{\hbar} \left(\frac{L}{ v_{\tqh}}+\frac{W}{ v_{\tsc}}\right)\right) \right]^{-1} .
\end{align}
This equation gives the complete expression for the chiral supercurrent carried by the chiral edge states for the geometry in Fig.~\ref{fig:1.pdf}, and is consistent with the result of Ref.~\onlinecite{Ma93} in the limit of $L \gg W$. 
In the high temperature limit, $\beta \hbar \ll ( L/ v_{\tqh}+ W/ v_{\tsc})$, this equation can be approximated as,
\begin{align}
I_\tsc \approx \frac{8e}{\beta\hbar}\sin \phi \, \sin^2 \left(\frac{\Delta W}{\hbar v_{\tsc}}\right) \textrm{exp}\left[-\frac{2\pi}{\beta\hbar} \left(\frac{L}{v_{\tqh}}+\frac{W}{ v_{\tsc}}\right)\right]. \label{ilt}
\end{align}

\section{Fraunhoffer periodicity} The current-phase relation can be obtained by including an external flux through the QH region. 
This can be incorporated by changing the gauge field $a(x)$ (Eq.~\eqref{gauge}) as,
\begin{align}
a(x)=\left\{\begin{array}{cl}
-\phi/2L+\phi_{e}/2L & \text{for } 0<x<L \\
\phantom{-} \phi/2L+\phi_{e}/2L & \text{for } L+W<x<2L+W \\
0 & \text{elsewhere } 
\end{array} \right. ,\label{gauge2}
\end{align}
where $\phi_{e}$ is the dimensionless external flux related to the actual flux $\phi_{\textrm {ext}}$ as  $\phi_{\textrm {ext}}=\phi_{e}\frac{\phi_0}{\pi}$. $\phi_0=h/2e$, is the superconducting flux quantum.

Including the flux $\phi_e$ in our calculation changes the supercurrent in Eq.~\eqref{fnl} to 
\begin{align} \label{flx}
I_{\tsc} (\phi_e)=-&\sum_{\omega_m}\frac{4e}{\beta\hbar}\sin \phi \, \sin^2 \left(\frac{\Delta W}{\hbar v_{\tsc}}\right) \nonumber \\
\times&\left[ (\cos \phi_e +\cos\phi)\cos \left(\frac{2 \Delta W}{\hbar v_{\tsc}}\right) +\cos \phi_e-\cos\phi \right.\nonumber \\
&~\left.-2\cosh\left(\frac{2\omega_m}{\hbar} \left(\frac{L}{ v_{\tqh}}+\frac{W}{ v_{\tsc}}\right)\right) \right]^{-1} .
\end{align}
We remark that in the parameter regime probed in the experiment the $\cosh\left(\frac{2\omega_m}{\hbar} \left(\frac{L}{ v_{\tqh}}+\frac{W}{ v_{\tsc}}\right)\right)$ term is by far the largest term of the denominator in the expression above. Moreover the $m=0,-1$ terms in the Matsubara frequency dominate. We can then approximate $I_{\tsc} $ as (Taylor expanding the denominator),
\begin{align}\label{cdc0}
	I_{\tsc}(\phi,\phi_e) \approx \Big( I_{\textrm{ch0}} + I_{\textrm{ch1}} \cos\phi_e \Big) \sin\phi.
\end{align}

\section{Comparison with the experimental results} Using experimental parameters of Ref.~\onlinecite{Amet966}, $\Delta=1.2 meV=13.9 K$, $W=2.4\mu m$, $L=0.3 \mu m$, $T=40 mK$, $B=1T$, cyclotron radius $r_c=25 nm$, and surface transparency $|t|\approx 0.7$, we can estimate edge velocities semi-classically (see Eq.~\eqref{Tapx2}) as $v_{\tqh}\approx 7.0\times 10^5 m/s$ and $v_{\tsc}\approx 3.9 \times 10^4 m/s$. 
Substituting these values into Eq.~\eqref{fnl} gives the magnitude of supercurrent $I_{\tsc}\approx 0.9 nA$, which is remarkably close to experimental value of $I_{\tsc} = 0.5 nA$. 
However, note that the exact value of this result should not be taken seriously since the exponential dependence of $I_{\tsc}$ on velocities ($v_{\tqh}$, $v_{\tsc}$) causes a large uncertainty in value of $I_{\tsc}$. 
Nonetheless, this result shows that a quantitative agreement in magnitude of the chiral supercurrent can be attained by making reasonable but critical assumptions about the SC/QH interface. 
Crucially, the exponential form of Eq.~\eqref{ilt} shows that the velocity renormalization and the surface transparency along the SC/QH interface play the main role in controlling the magnitude of supercurrent. 

From the order of magnitude difference between $v_\tqh$ and $v_\tsc$ in the exponential of Eq.~\eqref{ilt}, one can observe that geometrically the width of the superconducting contact ($W$) plays a crucial role in controlling the value of $I_{\tsc}$, whereas changing the length of the QH sample ($L$) does not cause much difference. 
This is consistent with the experimental observation of Ref.~\onlinecite{Amet966}. 
Moreover, and perhaps counter-intuitively, we find that decreasing the surface transparency of SC/QH interface $|t|$ can lead to an increase in magnitude of $I_{\tsc}$ by increasing $v_{\tsc}$. 
In the experiment, the $p$-doped regime has manifestly worse surface transparency (due to the PN junctions that are formed close to the contacts) and results in Ref.~\onlinecite{Amet966} actually shows larger value of $I_{\tsc}$ in that regime, supporting our theoretical conclusions. 

Let us now discuss the periodicity of the current-phase relation. The external flux $\phi_e$ dependence of the \emph{critical} chiral supercurrent $I_{\tsc}$ can be approximated as (from Eq.~\eqref{cdc0}), 
\begin{align}\label{cdc}
	I_{\tsc}^{\textrm c} (\phi_e) \equiv \textrm{max}_\phi  I_{\tsc}(\phi, \phi_e) \approx | I_{\textrm{ch0}} + I_{\textrm{ch1}} \cos\phi_e |,
\end{align}
where the $\phi_e$ independent term $I_{\textrm{ch0}}=0.9 \times 10^{-9} A$, and the $\phi_e$ dependent term $I_{\textrm{ch1}}=1.0 \times 10^{-11} A$, for the parameters we use. 
In apparent contradiction with the experiment (which is $ \phi_0=h/2e$ periodic), this expression suggests the supercurrent has a $2 \phi_0=h/e$ periodicity. 
However, it also shows that in the parameter regime of the experiment, external flux dependence of $I_{\tsc}$ is strongly suppressed in the sense that $I_{\textrm{ch1}}$ is almost two orders of magnitude smaller than $I_{\textrm{ch0}}$. Also the Fraunhoffer pattern of the chiral supercurrents do not form nodes as in conventional supercurrents. 

Given the strongly suppressed oscillations from the chiral supercurrent, one might wonder whether the experimentally observed period can be attributed to residual non-chiral supercurrent propagating through the system. Such non-chiral contributions can arise from, e.g., inhomogeneities in the confining potential near the edge. However, including such contributions (assuming they are smaller than $I_{\textrm{ch0}}$) does not change the periodicity. We are unable to explain the anomalous $\phi_0=h/2e$ periodicity observed in the experiment.

\section{Discussion and conclusion} In this paper we have studied the chiral supercurrent in a SC/QH/SC system for various system parameters. We have found that the finite junction transparency (consistent with normal state transport) and velocity renormalization along the SC contacts is crucial to obtain the correct order of magnitude of the supercurrent. In addition, we have found that in the high temperature limit, $\beta \hbar \ll ( L/ v_{\tqh}+ W/ v_{\tsc})$, both the flux averaged and flux dependent (giving $2\phi_0=h/e$ periodic Fraunhoffer pattern) chiral supercurrents go to zero exponentially with junction width with exponents $W\left[\frac{2\pi}{\beta\hbar} \left(\frac{L}{W v_{\tqh}}+\frac{1}{ v_{\tsc}}\right)\right]$ and $2W\left[\frac{2\pi}{\beta\hbar} \left(\frac{L}{W v_{\tqh}}+\frac{1}{ v_{\tsc}}\right)\right]$, respectively. 

We discussed the chiral nature of the supercurrent and showed that this ``chiral nature" can be used as a sharp definition for chiral supercurrents even in presence of the electron-electron interactions.

Acknowledgments: JS acknowledges support from the JQI-NSF-PFC, the National
Science Foundation NSF DMR-1555135 (CAREER) and the Sloan fellowship program.

\bibliographystyle{h-physrev}
\bibliography{library}

\begin{thebibliography}{10}
\bibitem{fnt}
Supercurrent is given by {$ -\frac{2e}{\hbar} \frac{\partial F}{\partial
  \phi}=\frac{2e}{\hbar\beta {Z}} \frac{\partial Z}{\partial \phi}$, where $Z$
  is the partition function: $Z=\int D(\bar{\psi},\psi) e^{-\int dx \sum_m
  [\bar{\psi}(x) (-i\omega_m+H(x))\psi (x) ] }$}.
\end{thebibliography}
\end{document}